\documentclass[12pt]{article}
\usepackage[dvips]{graphicx}

\def\v#1{{\bf#1}}
\def\be{\begin{equation}}
\def\ee{\end{equation}}
\def\bea{\begin{eqnarray}}
\def\eea{\end{eqnarray}}

\def\ie{{\it i.e.\,}}

\def\lcal{\mbox{$\cal L\,$}}
\def\ecal{\mbox{$\cal E\,$}}
\def\<{\langle}
\def\>{\rangle}

\begin{document}

\title{Propagators in the continuum limit: from molecules to scalar fields}

\author{ E. Sadurn\'i }

\date{Instituto de F\'{\i}sica\\
 Departamento de F\'isica Te\'orica \\
 Universidad Nacional Aut\'onoma de M\'exico\\
  Apartado Postal  20-364,
 01000 M\'exico D.F., M\'exico \\
 sadurni at fisica.unam.mx}

\maketitle

\begin{abstract}
The propagator of linear molecules whose constituents interact through oscillator potentials
can be obtained in a closed form for $N$ atoms as long as $N \leq 4$. We compute the propagator for
arbitrary $N$ in the approximation $N \gg 1$. Taking advantage of this result it is possible to analyze
the limit in which the molecule has an infinite number of constituents with infinitesimal length of sepparation,
corresponding to the quantization of a string, elastic rod or the second quantization of a Klein Gordon particle.
The evolution of some specific initial conditions is also studied, namely the time development of
states with minimal dispersion and the effect of sudden perturbations
on the vacuum of the scalar field theory. 
\end{abstract}




\vspace{2pc}


\maketitle

\section{Introduction}

\indent Transient phenomena in quantum mechanics have been of interest since the
developments in \cite{diff} a long time ago and others which are more recent \cite{kleber}, \cite{kramer}. The treatment of this subject demands a dynamical study of different quantum
systems corresponding to specific problems. In view of the simple theoretical results obtained for models involving oscillators in \cite{moshsadur}, there
arises naturally the question of their generalization to systems with infinite degrees of freedom.

In many textbooks \cite{goldstein} we find treatments of continuous media as the limit of interacting discrete systems
in the context of classical mechanics. However, it is possible to exploit this analogy in the quantum picture
through the Feynman formulation of the propagator \cite{Feynman}, connecting the latter with classical lagrangians. In
this fashion one may study discretized versions of second order field equations and quantize them before
taking the continuum limit. The inverse order of this limits makes it possible to perform certain calculations,
while when taken in the usual order gives rise to path integrals of field configurations. The latter may be
difficult to handle or may even lack of an appropriate mathematical definition.

Our prototype discrete system will be the linear molecule and we
proceed to compute its propagator in section 2. In section 3 the continuum limit of our system is taken
calculating thereby the propagator for strings and second quantized scalar fields. In sections 4 and 5 we apply
our results to certain initial conditions, including sudden external interactions.

\section{Propagator of the linear molecule}

We start with a lagrangian in $1+1$ dimensions of a chain of $N$ particles of equal masses $m$ connected
by equal springs of constant $k=m \omega^2$

\bea
\lcal= \frac{m}{2} \sum^{N}_{i=1}\left( \dot x_{i}^2 - \omega^2 (x_{i}-x_{i+1})^2  \right) \equiv \sum^{N}_{i=1} \lcal_i
\label{e1}
\eea
It is costumary to find the orthogonal transformation which sepparates the problem in $N$ independent
lagrangians (normal modes) and we will do so in the following. Letting $(\v x)_i = x_i$ and

\bea
\v V = \left( \begin{array}{c c c c c c c c} 1 & -1 & 0 & 0 & ... & & & 0 \\ -1 & 2 & -1 & 0 & ... & & & 0 \\ 0 & -1 & 2 & -1 & ... & & & 0 \\ .& & & & & & & \\ .& & & & & & & \\ 0 & ... & & & & -1 & 2 & -1 \\ 0 & ... & & & & 0 & -1 & 1 \end{array} \right)_{N \times N}
\label{e2}
\eea
our lagrangian becomes

\bea
\lcal= \frac{m}{2} \left( \dot \v x^2 - \omega^2 \v x^T \v V \v x \right)
\label{e3}
\eea
and it is left to diagonalize $\v V$. It can be shown (see Appendix) that the eigenvalues of $\v V$ are given by

\bea
\lambda_n = 4 \sin^2{ \left( \frac{n \pi}{2(N-1)} \right) } + O(1/N), \qquad 0 \leq n \leq N-1
\label{e4}
\eea
which is particularly useful for large $N$. We can see that $N \sim 10$ is already a good approximation as can be
checked numerically. We resort to this type of formula since it allows to manipulate our
expressions simbolically though numerical methods to compute $\lambda_n$ exist. Notice that in the limit
$N \rightarrow \infty $ the $\lambda$'s (frequencies of the normal modes) populate densely the interval
$[0,4]$ distributed by a trigonometric law. It is straightforward then to write

\bea
\lcal= \frac{m}{2} \left( \dot \v y^2 - \omega^2 \v y^T \v D \v y \right)
\label{e5}
\eea
where $\v y = \v O \v x$, $\v O$ being the orthogonal transformation corresponding to the diagonalization of
$\v V$ into $\v D= \rm diag \ (\lambda_0,...,\lambda_{N-1}) $. The propagator is written as \cite{grosche}

\bea
 \nonumber K(\v y,\v y';t)=\prod^{N-1}_{n=0} \sqrt{\frac{m \omega \lambda_n^{1/2}}{2\pi i \hbar \sin{ \left( \omega \lambda_n^{1/2} t \right) }}} \times \\ \times \exp{ \left[ \frac{i m \omega \lambda_n^{1/2}}{2 \hbar \sin{ \left( \omega \lambda_n^{1/2} t \right) } }
\left( \cos{ \left( \omega \lambda_n^{1/2} t \right) }(y_{n}^2 + {y'}_{n}^2)-2 y_n {y'}_n \right) \right]}
\label{e6}
\eea
and this expression is exact as long as the $\lambda$'s contain corrections $O(1/N)$, but these do not alter
the functional form of (\ref{e6}). Notice that the factor associated to $n=0$ corresponds to $\lambda_{0}=0$
and can be written as

\bea
K_{0}(\v y_{0},\v {y'}_{0};t) = \sqrt{\frac{m}{2\pi i \hbar t}} \exp{ \left[ \frac{i m }{2 \hbar t}\left( y_{0} - {y'}_{0} \right)^2 \right] }
\label{e6bis}
\eea
which is the propagator of the free particle, finding thereby that $y_{0}$ is the coordinate
of the center of mass. The meaningful degrees of freedom are therefore indicated by $\v y$,
but we can at any moment employ 
the inverse transformation $\v x = \v O^T \v y $ since $\v O$ can be written explicitly (see Appendix).

\section{The continuum limit of the linear molecule propagator}

Here we will consider the continuum limit as that indicated in \cite{goldstein} or \cite{kaku}.
First of all let us recall that the lagrangian (\ref{e1}) can be multiplied by a parameter $a$ with the
dimensions of length so that

\bea
\lcal=  \sum^{N}_{i=1} a \left( \frac{m}{2a} \dot x_{i}^2 - \frac{m \omega^2 a}{2} (\frac{x_{i}-x_{i+1}}{a})^2  \right) \\
\begin{array}{c} \longrightarrow \\ {_{N \rightarrow \infty, \quad a \rightarrow 0}} \end{array} \int^{L}_{0} d\xi \frac{\mu}{2}(\partial_t \phi)^2 - \frac{\mu \Omega^2}{2}(\partial_{\xi} \phi)^2
\label{e7}
\eea
as long as $x_i(t) \rightarrow \phi(\xi,t)$, $a=\frac{L}{N} \rightarrow d\xi$, $\omega a \rightarrow \Omega$.
The constant $L$ is interpreted as the length of the resulting string and
$\Omega$ is such that $\mu \Omega^2 = Y$, \ie Young's module.

For the propagator limit we use Feynman's prescription

\bea
\nonumber K(\v x,\v x';t)=\left( \prod^{N}_{i=1} \int^{x_i(t)=x_f}_{x_i(0)=x_0} Dx_i(\cdot) \right) \exp{ \left( -\frac{i}{\hbar}\int^{t}_{0}d\tau \sum^{N}_{i=1} \lcal_i \right) } \\
\begin{array}{c} \longrightarrow \\ _{N \rightarrow \infty} \end{array} \int^{\phi(x_f,t)}_{\phi(x_0,0)} D\phi(\cdot) \exp{ \left( \frac{-i}{\hbar}\int^{t}_{0}\int^{L}_{0}d\tau d\xi \frac{\mu}{2}(\partial_t \phi)^2 - \frac{\mu \omega^2}{2}(\partial_{\xi} \phi)^2 \right) }
\label{e11}
\eea
which is the propagator of a string, elastic rod or a massless scalar relativistic particle when the classical
length $L \rightarrow \infty$ and the identification $\Omega \leftrightarrow c$ takes place.

Though we are quite familiar with this form of the limit, in normal coordinates $\v y$ some subtleties
arise. First of all, we notice that the $N \rightarrow \infty $ limit of lagrangian (\ref{e5}) is equal to the limit
of the following lagrangian

\be
\sum_{j=1}^{N} a \left( \frac{\mu}{2} \dot y_j^2 - \frac{\mu \Omega^2}{2} (\lambda_j / a^2) y_j^2 \right)
\label{e11.1}
\ee
where we can observe that the eigenvalues are now $\lambda_j / a^2$. Since 

\be
\frac{\lambda_j}{a^2}= \frac{4N^2}{L^2} \sin^2 \left( {\frac{j \pi}{2(N-1)}} \right) + O(1/N) \begin{array}{c} \longrightarrow \\ {_{N \rightarrow \infty}} \end{array} \left( \frac{j \pi}{L} \right)^2 \equiv \Lambda_j
\label{e11.2}
\ee
we see that these eigenvalues are distributed discretely for finite $j$ by a square law, just as
the energy spectrum of a string with given initial and final conditions at a fixed time. Therefore, the limit
of (\ref{e11.1}) is not an integral, but a series and it will make sense if we set $\sqrt{a} y_j \rightarrow \eta_j$,
\ie if fields are rescaled. To this effect we should also rescale the orthogonal operator $\v O$ in the
following way

\be
\frac{1}{\sqrt{a}} O_{ij} \begin{array}{c} \longrightarrow \\ {_{N \rightarrow \infty}} \end{array} O_{j}(\xi)
\label{e11.3}
\ee
so that the transformations between fields become

\bea
 \eta_j = \int_{0}^{L} d \xi O_j(\xi) \phi(\xi), \qquad \phi(\xi) = \sum_{j} O_j(\xi) \eta_j  \nonumber \\
\int_{0}^{L} O_j(\xi)O_l(\xi) = \delta_{jl}, \qquad \sum_{j} O_j(\xi) O_j(\xi') = \delta(\xi-\xi')
\label{e11.4}
\eea
where $ a^{-1} \delta_{ii'} \rightarrow \delta(\xi-\xi')$. With all these considerations, lagrangian (\ref{e5})
posseses the limit

\be
\lcal \rightarrow \sum_{j=0}^{\infty} \frac{\mu}{2} \dot \eta_j^2 - \frac{\mu \Omega^2}{2} (j \pi / L)^2 \eta_j^2
\label{e11.5}
\ee
whose discrete form stems from the discreteness of the eigenvalues $\Lambda_j$ as we have proven. 

The propagator related to lagrangian (\ref{e11.5}) can be written directly or it can be computed as the limit
$N \rightarrow \infty$ of propagator (\ref{e6}) but taking into account that the prefactor is proportional to
$m^{N/2} \sim a^{N/2} \mu^{N/2} $. Thus, for the fields $\eta_j$ the propagator is multiplied by the jacobian
$a^{-N/2}$, cancelling the scale factor emerging from the mass. The resulting expression is

\bea
 \nonumber K_{{\rm string\ }}(\eta,\eta';t)= \\ f(t) \exp \left( \frac{i \mu \Omega}{2 \hbar} \sum_{j=0}^{\infty} \frac{j \pi}{L \sin (\Omega j \pi t/ L)} \left[ \left( \eta_j^2 + \eta_j^{'2} \right) \cos (\Omega j \pi t/ L)-2 \eta_j \eta'_j \right] \right)
\label{e11.6}
\eea
with

\be
f(t) \equiv \prod_{j=0}^{\infty} \frac{\mu \Omega j \pi}{2 \pi i \hbar L \sin (\omega j \pi t/L)}
\label{e11.7}
\ee

The definition for $f$ deserves some comments. The convergence of this infinite product can be spoiled
by the presence of $\mu$ and $\Omega$, leading to $f= \infty, 1$ or $0$ for different values of these parameters.
Nevertheless the propagator itself obeys a normalization condition even when the prefactor is a divergent or a 
vanishing quantity. This stems from the fact that $K$ is the representation of a unitary operator and preserves
norm. The limit of propagators should be taken under the integral $\int D \phi \left[ \cdot \right]$ or its discrete version
from which it is easy to see that a rescaling of fields $\bar \phi = \sqrt{\frac{\mu \Omega}{2 \pi \hbar}} \phi$,
$\bar \eta = \sqrt{\frac{\mu \Omega}{2 \pi \hbar}} \eta$ gets rid of the problem. Using dimensionless fields $\bar \eta$
is the price to be paid. With all this the propagator becomes

\be
 K_{{\rm string\ }}(\bar \eta,\bar \eta';t)=\bar f(t) \exp \left( i \sum_{j=0}^{\infty} \frac{j \pi^2}{L \sin (\Omega j \pi t/ L)} \left[ \left( \bar \eta_j^2 + \bar \eta_j^{'2} \right) \cos (\Omega j \pi t/ L)-2 \bar \eta_j \bar \eta'_j \right] \right)
\label{e11.8}
\ee
with

\be
\bar f(t) \equiv \prod_{j=0}^{\infty} \frac{ j \pi / L}{i \sin (\omega j \pi t/L)} = \sqrt{\det \left[ i\left( V+i \epsilon \right)\left(\sin(V+i\epsilon)\right)^{-1} \right]}
\label{e11.9}
\ee
$V \equiv V_{\infty}$ and $\epsilon$ arbitrarily small. 
Let us now write the propagator of a second-quantised Klein Gordon particle of mass $M$. First of all
we need to add a mass term
$\frac{m M^2c^2}{2 \hbar}\v x \cdot \v x$ to lagrangian (\ref{e1}). Since this term is invariant under the
orthogonal transformation with associated matrix $\v O$, it is sufficient to replace $\omega$ in (\ref{e6})
by $\omega_j = \sqrt{\omega^2 + \frac{M^2c^4}{\hbar^2 \lambda_j}}$. As we have indicated, we must set $\Omega = c$. Once this is done we need to extend the molecule
to the whole real line by taking $L \rightarrow \infty$. We identify the continuous wave number as
$j \pi /L \rightarrow k$, giving a differential $\pi/L \rightarrow dk$. The fields $\bar \eta$ must be
rescaled again in order to have the limit $\sqrt{L} \bar \eta_j \rightarrow \eta(k)$. The
orthogonal transformation obeys $\sqrt{L}O_j(\xi) \rightarrow O(\xi,k)$ in order to have field transformations in full integral form, \ie 

\bea
 \nonumber \eta(k) = \int_{-\infty}^{\infty} d \xi O(k,\xi) \phi(\xi), \qquad \phi(\xi) = \int_{-\infty}^{\infty} d k O(\xi,k) \eta(k) \\
 \int_{-\infty}^{\infty} d\xi O(k,\xi)O(k',\xi) = \delta(k-k'), \qquad \int_{-\infty}^{\infty} dk O(\xi,k) O(\xi',k) = \delta(\xi-\xi')
\label{e11.10}
\eea
where $L\delta_{jj'} \rightarrow \delta(k-k')$. Meanwhile, the lagrangian (\ref{e11.5}) with
a mass term has the following limit when $L \rightarrow \infty$

\be
\int_{-\infty}^{\infty} dk | \dot \eta(k) |^2 - (E_k / \hbar)^2 | \eta(k)|^2
\label{e11.11}
\ee
where we have computed 

\be
\hbar \lambda_j^{1/2} \omega_j \begin{array}{c} \longrightarrow  \\ {_{L \rightarrow \infty}}  \end{array} \sqrt{\hbar^2 c^2 k^2 + M^2 c^4} \equiv E_k
\label{e11.12}
\ee

From the equations of motion induced by this lagrangian we see that spatial derivatives become factors of $k$, inferring thus that $O(k,\xi)$ is the Fourier kernel.
Finally the sought propagator is given by

\bea
 \nonumber K_{KG}\left[ \eta, \eta', t \right]= \\  g(t) \exp \left[ \int_{-\infty}^{\infty} dk \frac{i E_k}{\hbar^2 k \sin \left( \frac{E_k t}{\hbar} \right) } \left[ \cos \left( \frac{E_k t}{\hbar} \right) \left( | \eta(k) |^2 + |\eta'(k)|^2 \right) - 2 \eta(k) \eta'(k) \right]  \right]
\label{e11.13}
\eea
with

\be
g(t) \equiv \prod_{k \in \v R} \sqrt{\frac{E_k}{i\hbar \sin \left( \frac{E_k t}{\hbar} \right)}}
\label{e11.14}
\ee
The prefactor is given again in terms of an infinite product and it is left left indicated without discussing its convergence. At any rate, it seems more suitable to work with the molecule model and its propagator (\ref{e6}) or its { \it massive \ } version in order to study dynamical effects on strings or particle fields. Any meaningful time dependent quantity of these systems will be treated discretely and limits will be computed at the end of calculations.

\section{Minimally localized molecule and its continuum limits}

At this point we are ready to apply propagator (\ref{e6}) to some initial conditions. Consider the linear
molecule of $N$ atoms to be in a state which is minimally dispersed, \ie a gaussian distribution in space
for each atom:

\bea
\psi(\v x',0)= \left( \frac{1}{2 \pi \sigma^2} \right)^{N/4} \exp{ \left[ -\frac{1}{4\sigma^2}\sum^{N-1}_{i=0} x'^2_i \right] }
\label{e17}
\eea
where the width $\sigma$ is the same for all components. This wave function has the property

\bea
\psi(\v x',0)=\psi(\v O \v x',0)=\psi(\v y',0). 
\label{e18}
\eea
and it describes a localization in phase space with minimal uncertainty.
In the continuum limit $N \rightarrow \infty$, initial condition (\ref{e17}) becomes

\bea
\psi \left[ \phi, 0 \right] &=& \lim_{N \rightarrow \infty} \left( \frac{1}{2 \pi \sigma^2} \right)^{N/4} \exp \left( -\frac{1}{4 s^2} \int_{0}^{L} d\xi \phi^2(\xi) \right) \nonumber \\
&=& \lim_{N \rightarrow \infty} \left( \frac{1}{2 \pi \sigma^2} \right)^{N/4} \exp \left( -\frac{1}{4 s^2} \sum_{j=0}^{\infty} \eta_j^2 \right)
\label{e18.1}
\eea
where $\sigma^2 a \rightarrow s^2$. Similar remarks to those following (\ref{e11.7}) hold
here. Despite the limit of the factor preceding the exponential may be vanishing or divergent, functional (\ref{e18.1}) is normalized and for all $t$ we have
$\int D \phi |\psi \left[\phi,t \right]|^2 = 1$. Thus, functional (\ref{e18.1}) is a probability amplitude depending on field configurations. In order to obtain a normalized functional in normal coordinates we need though to use the jacobian $a^{-N/2}$ to write the initial condition as

\bea
\tilde \psi \left[ \eta, 0 \right] = \left[ \lim_{N \rightarrow \infty} \left( \frac{1}{2 \pi s^2} \right)^{N/4} \right] \exp \left( -\frac{1}{4 s^2} \sum_{j=0}^{\infty} \eta_j^2 \right)
\label{e18.2}
\eea
so that 

\be
\int \prod_j d \eta_j |\tilde \psi \left[ \eta, 0 \right] |^2 = 1
\label{e18.3}
\ee
Now we apply (\ref{e6}) to (\ref{e18}) by integrating over $\v y'$ to obtain

\bea
\psi(\v y,t)= \left( \prod^{N-1}_{j=0} \left( \frac{1}{2 \pi \sigma^2_j(t)} \right)^{1/4} \right) \exp{ \left[ -\sum^{N-1}_{j=0}\frac{y^2_i}{4\sigma^2_j(t)} \right] } e^{i \Delta(t)}
\label{e19}
\eea
with

\bea
\sigma_j(t)=\sigma \sqrt{1+ \left( \frac{\hbar^2}{2 \omega^2 m^2 \sigma^4 \lambda_j}-1 \right) \sin^2{ \left( \omega \lambda_j^{1/2} t \right) }}
\label{e20}
\eea
and $\Delta(t)$ a phase which is irrelevant to the probability density:

\bea
|\psi(\v y,t)|^2= \left( \prod^{N-1}_{i=0} \left( \frac{1}{2 \pi \sigma_i^2(t)} \right)^{1/2} \right) \exp{ \left[ -\sum^{N-1}_{i=0}\frac{y^2_i}{2\sigma^2_i(t)} \right] } 
\label{e21}
\eea
We observe therefore an oscillatory behaviour of the density widths and amplitudes depending on  the corresponding frequencies $\omega \lambda_j^{1/2}$. The center
of mass ($\lambda_{0} = 0$) distribution exhibits disispation as expected. In the continuum limit (\ref{e20}) yields

\be
\sigma_j(t) \begin{array}{c} \longrightarrow \\ {_{N \rightarrow \infty}} \end{array} s_j(t)= s \sqrt{1+ \left( \frac{\hbar^2 L^2}{2 \Omega^2 \mu^2 s^4 j^2 \pi^2}-1 \right) \sin^2{ \left( \Omega t j \pi / L \right) }}
\label{e24}
\ee
whose maximum value decreases with $j$. For the massive Klein Gordon particle we
find a different limit. The discrete case yields

\bea
\sigma_j(t)=\sigma \sqrt{1+ \left( \frac{\hbar^2}{2 \omega_j^2 m^2 \sigma^4 \lambda_j}-1 \right) \sin^2{ \left( \omega_j \lambda_j^{1/2} t \right) }}
\label{e25}
\eea
Taking $N \rightarrow \infty$, $L \rightarrow \infty$, initial condition (\ref{e18.2}) becomes

\be
\tilde \psi \left[ \eta, 0 \right] = \left[ \lim_{N \rightarrow \infty} \left( \frac{1}{2 \pi \bar s^2} \right)^{N/4} \right] \exp \left( -\frac{1}{4 \bar s^2 } \int_{-\infty}^{\infty} dk \eta^2(k) \right)
\label{e25.1}
\ee
after rescaling widths $\bar s = \sqrt{\frac{\mu \Omega}{ 2 \hbar}} s $. The wave functional for fields $\eta(k)$ at time $t$ is now 

\be
\tilde \psi \left[ \eta, t \right] = \left[ \lim_{N \rightarrow \infty} \left( 
\frac{1}{2 \pi s_k^2(t)} \right) \right] \exp \left( - \int_{-\infty}^{\infty} dk \frac{ |\eta(k)|^2}{4 s_k^2(t)} \right)
\label{e25.2}
\ee
with

\be
s_k^2(t) = \bar s \sqrt{1 + \left( \frac{2 \hbar^2 c^2}{E_k^2 \bar s^4}-1\right) \sin \left( \frac{E_k t}{ \hbar } \right) }
\label{e25.3}
\ee
This quantity is again oscillatory but with a frequency given by the relativistic formula of kinetic energy. The most probable field configuration remains $\eta \equiv 0$, but the field
localization width can spread to a factor $(\sqrt{2} \hbar c)/( E_k \bar s^2)$.

To illustrate the significance of this result let us
compute the average number of quanta for each mode using states $\psi(\v y,t)$. In the discrete case we have

\bea
\< N_j \> &=& \< \quad (\hbar \omega \lambda_j^{1/2})^{-1}  H_{ \rm oscillator \ } \left(-i\frac{\partial}{\partial y_j},y_j,\omega \lambda_j^{1/2} \right)-1/2 \quad \> \\
 &=& \frac{m \omega}{2 \hbar} \sigma_j^2(t) \lambda_j^{1/2} + \frac{3\hbar}{4m \omega}(\sigma_j^2(t) \lambda_j^{1/2})^{-1}-\frac{1}{2}
\label{e27}
\eea
Taking the continuum limit $N \rightarrow \infty$ the average number results in

\bea
\< N_j \> = \frac{\mu \Omega}{2 \hbar} s_j^2(t) \left( \frac{j \pi}{L} \right) + \frac{3\hbar}{4 \mu \Omega}(s_j^2(t) \left( \frac{j \pi}{L} \right))^{-1}-\frac{1}{2}
\label{e27bis}
\eea
with $s_j(t)$ given by (\ref{e24}). For the massive relativistic particle we have
to replace the frequencies by $\omega_j$ and take $L \rightarrow \infty$. The expectation values computed by integrating with respect to fields $\eta(k)$ are now multiplied by the squared jacobian $1/L \rightarrow dk / \pi$, thus we get

\bea
\frac{1}{L} \< N_j \> \begin{array}{c} \longrightarrow \\ {_{L \rightarrow \infty}} \end{array} \left[ \frac{ E_k }{ \hbar} s_k^2(t) + \frac{3\hbar}{8 E_k}(s_k^2(t))^{-1}-\frac{1}{2} \right] \frac{dk}{\pi} \equiv \nu_k(t) dk
\label{e27bisbis}
\eea
where we have obtained a number density $\nu_k(t)$ in momentum space as a function of time.

\begin{figure}[!h]
\begin{center}
\includegraphics[width=12cm]{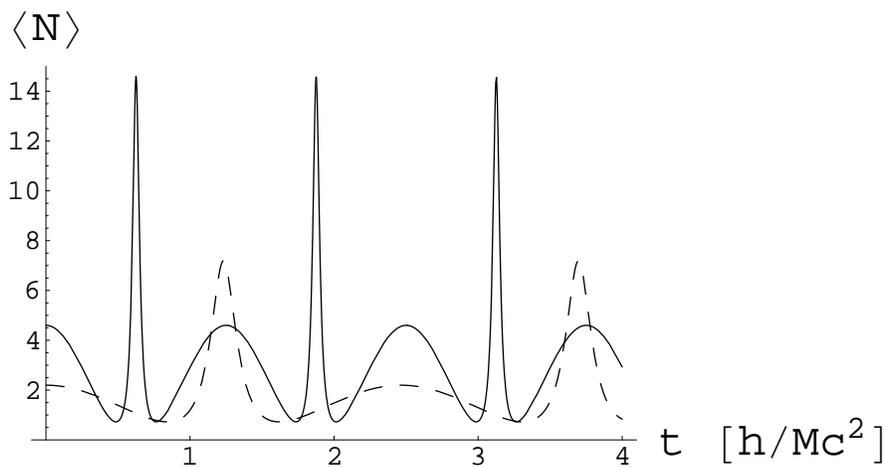}
\end{center}
\caption{Average number density of particles as a function of $t$ for $k=\frac{10 Mc}{\hbar}$ (Solid line) and $k=\frac{5 Mc}{\hbar}$ (Dashed line).}
\label{figure1}
\end{figure}
An oscillatory density number is not a surprising result when viewed as the evolution of harmonic oscillator
number of quanta. However, it is interesting to notice that such density number corresponds
to a second-quantised { \it free \ } field which has beeen set initially in the most
localized configuration in phase space (gaussian distributions).

\section{The string under a sudden electrostatic field and the relativistic particle with a source}

Now we may think of another example which involves a scalar field and a sudden interaction with a source which is equivalent to a charged string between the plates of a condensator in the discrete case. Consider a system whose
lagrangian is given by

\bea
\lcal' = \lcal + \theta(t) \v \ecal \cdot \v x
\label{e28}
\eea
where $\lcal$ is given by (\ref{e1}), $\theta(t)$ is the step function and $\v \ecal$ denotes some electrostatic
field multiplied by the charge of the $i$th particle in the molecule, \ie $\ecal_i=q_i E_i$ (remember that
this is a one dimensional example and vectors here stand for the $N$ components of our system). If we set
$\frac{q_i}{a} \rightarrow \rho(\xi)$ as the linear charge density and $\frac{q_i E_i}{a} \rightarrow \rho(\xi)E(\xi) \equiv \ecal(\xi)$, then the limit of (\ref{e28}) is well
known

\bea
\lcal'  \begin{array}{c} \longrightarrow \\ _{N \rightarrow \infty} \end{array} \int^{L}_{0} d\xi \left( \frac{\mu}{2}(\partial_t \phi(\xi,t))^2 - \frac{\mu \Omega^2}{2} (\partial_{\xi} \phi(\xi,t))^2 + \theta(t) \ecal(\xi) \phi(\xi,t) \right)
\label{e29}
\eea
and leads to equations of motion \cite{kaku}

\bea
\partial_t^2 \phi - \Omega^2 \partial_{\xi}^2 \phi = J(\xi), \quad t>0
\label{e30}
\eea
where we have defined a current $J(\xi) = \frac{2}{\mu} \ecal(\xi)$. But before taking the continuum limit,
let us assume that for $t<0$ the discrete system is in its ground state, \ie

\bea
\psi(\v y,0)=\prod^{N-1}_{i=0} \left( \left( \frac{\hbar}{\pi m \omega \lambda_i^{1/2}} \right)^{1/4} \exp{ \left[ - \frac{m \omega \lambda_i^{1/2}}{2 \hbar} y_i^2 \right] } \right) = |0,...,0 \>
\label{e31}
\eea
given, of course, in normal coordinates. The source term in (\ref{e28}) can be written as

\bea
\v \ecal \cdot \v x = \v \ecal' \cdot \v y, \qquad \v \ecal'= \v O \v \ecal 
\label{e32}
\eea
With this orthogonal transformation of the external field and for positive times we can immediately write

\bea
\lcal'(\dot \v y,\v y) = \lcal(\dot \v z,\v z) + \sum^{N-1}_{i=0} \frac{\ecal'^2_i}{2m \omega_i^2 \lambda_i} 
\label{e33}
\eea
where $z_i \equiv y_i - \frac{\ecal'_i}{m \omega_i^2 \lambda_i}$. In the $N \rightarrow \infty$ limit we must set $\ecal'_j/\sqrt{a} \rightarrow F_j$ so that

\bea
 F_j = \int_{0}^{L} d\xi O_j(\xi) \ecal(\xi) 
\label{e33.bis}
\eea
The propagator related to lagrangian (\ref{e33}) has the form
(\ref{e6}) but evaluated in $\v z$ and with an additional (but irrelevant) phase factor containing the field
energy.

It is quite easy to apply such propagator to the vacuum state (\ref{e31}) and we write the resulting probability
density computed in \cite{moshsadur} for each mode

\bea
|\psi_j(\v y,t)|^2= \left( \frac{\pi \hbar}{m \omega \lambda_i^{1/2} } \right)^{1/2} \exp{ \left[-\frac{m \omega \lambda_j^{1/2}}{\hbar}\left( y_j + a_j(t) \right)^2 \right] }
\label{e34}
\eea
where we can see a time dependent average of the distribution given by

\bea
a_j(t)= \frac{\ecal'_j \sin^2(\omega \lambda_j^{1/2} t/2) } { m \omega^2 \lambda_j }.
\label{e35}
\eea
The continuum limit yields

\bea
a_j(t) / \sqrt{a} \longrightarrow  \frac{F_j \sin^2(\Omega t j \pi /(2L))}{\mu \Omega^2 
(j \pi / L)^2} \equiv \alpha_j(t)
\label{e36}
\eea
For the Klein Gordon Field we again rescale fields and set, therefore, $\sqrt{L} F_j \rightarrow F(k)$ so that $F(k)$ can be the Fourier transform of $\ecal(\xi)$ when $L \rightarrow \infty$. The average
field configuration (\ref{e36}) in the massless case yields

\be
\sqrt{\frac{\mu \Omega L}{2 \hbar}} \alpha_j (t) \begin{array}{c} \longrightarrow  \\ {_{L \rightarrow \infty}}  \end{array} \alpha(k,t) = I(k) \frac{ \sin^2 \left( ctk \right) }{2 c^2 k}
\label{e36.1}
\ee
where $I(k)$ is the Fourier transform of $J(\xi)$. As before, the massive case is obtained by replacing the frequencies and the result is

\bea
\alpha(k,t) = \frac{\hbar^2 I(k) \sin^2 \left( \frac{E_k t}{2 \hbar} \right)}{2 E_k^2}
\label{e36.2}
\eea
The average number taken between states (\ref{e34}) yields, in the discrete case, a number density in momentum space after taking the limits $N \rightarrow \infty$, $L \rightarrow \infty$. Proceeding analogously to the derivation of (\ref{e27bisbis}) we arrive at

\bea
\frac{1}{L}\< N_j \> \begin{array}{c} \longrightarrow \\ {_{L \rightarrow \infty}}  \end{array} \frac{\hbar^3 |I(k)|^2 \sin^4 \left( \frac{E_k t}{2 \hbar}\right)}{4 E_k^2} dk \equiv \nu_k(t) dk
\label{e36.3}
\eea
In this result we observe that the Fourier transform of the current ($I(k)$) determines whether the number density $\nu$ increases with energy or not, depending on the explicit form of
$J(\xi)$. Another unexpected feature is that even when the source is constant
in time (say $t>0$),
the number oscillates with a frequency depending on the relativistic energy of the produced particles.

\section{Conclusions}

Discrete systems are useful to study quantum dynamical problems whose continuum
limits may be difficult to formulate in quantum field theory. We have outlined a method by which
we can obtain propagators (\ref{e6},\ref{e11.8},\ref{e11.13}), functionals (\ref{e19},\ref{e21}, \ref{e34}) and
physical quantities (\ref{e27bis},\ref{e27bisbis},\ref{e36.3}) by developing some simple examples:
molecules, strings (or elastic rods) and Klein Gordon particles in second quantization.
The method allows interactions in a non perturbative regime, but their form is kept simple
so that lagrangians remain quadratic in their discrete version.

\appendix
\section*{Appendix}

\setcounter{section}{1}

Now we proceed to derive expression (\ref{e4}). Consider the matrix $\v V_N$ as in (\ref{e2}) and let

\bea
\v M_N = \left( \begin{array}{c c c c c c c c} 1 & -1 & 0 & 0 & ... & & & 0 \\ -1 & 2 & -1 & 0 & ... & & & 0 \\ 0 & -1 & 2 & -1 & ... & & & 0 \\ .& & & & & & & \\ .& & & & & & & \\ 0 & ... & & & & -1 & 2 & -1 \\ 0 & ... & & & & 0 & -1 & 2 \end{array} \right)_{N \times N}
\label{a1}
\eea
Define also the polynomials

\bea
\phi_N(\lambda)= |\v V_N - \lambda \v I_N|, \qquad \chi_N(\lambda)= |\v M_N - \lambda \v I_N|
\label{a2}
\eea
It can be easily verified that the following recursion relations hold

\bea
\left(\begin{array}{c} \phi_N(\lambda) \\ \chi_N(\lambda) \end{array} \right)=\left(\begin{array}{c c} 1 & -\lambda \\ 1 & 1-\lambda \end{array} \right)\left(\begin{array}{c} \phi_{N-1}(\lambda) \\ \chi_{N-1}(\lambda)  \end{array} \right)
\label{a3}
\eea
The matrix appearing in the RHS of (\ref{a3}) can be diagonalized and then the recursion is solved by
computing the powers of that matrix. Let $a= \frac{1}{2}(\lambda+\sqrt{\lambda(\lambda-4)})$ and choose
as initial conditions $\phi_2, \chi_2$ which are given by

\bea
\phi_2(\lambda)= \lambda^2-2\lambda, \qquad \chi_2(\lambda)= \lambda^2-3\lambda + 1
\label{a4}
\eea
Following the procedure indicated in the last paragraph by finding the appropriate similarity transformation
we can write (\ref{a3}) as

\bea
 \nonumber \left(\begin{array}{c} \phi_N(\lambda) \\ \chi_N(\lambda) \end{array} \right)= (\lambda - a^2)^{-1} \left(\begin{array}{c c} 1 & 1 \\ \frac{a}{\lambda} & \frac{1}{\lambda} \end{array} \right)  \left(\begin{array}{c c} (1-a)^{N-2} & 0 \\ 0 & (1-\frac{\lambda}{a})^{N-2} \end{array} \right) \times \\ \times  \left(\begin{array}{c c} \lambda & -a \lambda \\ -a^2 & a \lambda \end{array} \right)  \left(\begin{array}{c} \phi_{2}(\lambda) \\ \chi_{2}(\lambda)  \end{array} \right)
\label{a5}
\eea
from which the first component is used to write the secular equation $\phi_N(\lambda)=0$. We notice
that the reality of $\lambda$ implies $a=e^{i\alpha}$ and therefore $\lambda= 4 \cos^2{(\alpha/2)}$.
After some algebraic steps we find that the secular equation is equivalent to

\bea
 (\phi_2(\lambda)-2\chi_2(\lambda)) \sin{\left( (N-2) \alpha \right)} \cos{(\alpha/2)} - \phi_2(\lambda) \cos{\left( (N-2) \alpha \right)} \sin{(\alpha/2)} = 0
\label{a6}
\eea
Now we can estimate the solutions of (\ref{a6}) by making the Ansatz

\bea
\alpha = \frac{N-1-n}{N-1} \pi 
\label{a6.1}
\eea
for finite $n$. Replacing back in (\ref{a6}) and neglecting terms $O(1/N)$ we see
that the secular equation is solved. Therefore

\bea
\lambda = 4 \sin^2 \left( \frac{n \pi}{2(N-1)} \right) + O(1/N) 
\label{a6.2}
\eea
We can also write the equations for the eigenvectors in recurrence form and these can be solved in terms of the
eigenvalues given above.

\end{document}